\documentclass[preprint,aps]{revtex4}
\usepackage{graphicx,epsfig}
\begin{document} 
\def\rez{$^{1}$}
\def\gsi{$^{2}$}
\def\hei{$^{3}$}
\def\dub{$^{4}$}
\def\wei{$^{5}$}
\def\sun{$^{6}$}
\def\cer{$^{7}$}
\def\bnl{$^{8}$}
\def\mpi{$^{9}$}

\title{Enhanced Production of Low-Mass Electron-Positron Pairs in\\
40 AGeV Pb-Au Collisions at the CERN SPS\footnote{Doctoral Thesis of S. Damjanovi\'c, University of Heidelberg (2002)}}
\author{D.~Adamov\'a\rez, 
G.~Agakichiev\gsi, 
H.~Appelsh\"auser\hei, 
V.~Belaga\dub, 
P.~Braun-Munzinger\gsi, 
A.~Cherlin\wei, 
S.~Damjanovi\'c\hei,
T.~Dietel\hei, 
L.~Dietrich\hei, 
A.~Drees\sun, 
S.\,I.~Esumi\hei, 
K.~Filimonov\hei, 
K.~Fomenko\dub,
Z.~Fraenkel\wei, 
C.~Garabatos\gsi, 
P.~Gl\"assel\hei, 
G.~Hering\gsi, 
J.~Holeczek\gsi, 
V.~Kushpil\rez, 
B.~Lenkeit\cer,
A.~Maas\gsi, 
A.~Mar\'{\i}n\gsi, 
J.~Milo\v{s}evi\'c\hei,
A.~Milov\wei, 
D.~Mi\'skowiec\gsi, 
Yu.~Panebrattsev\dub, 
O.~Petchenova\dub, 
V.~Petr\'a\v{c}ek\hei, 
A.~Pfeiffer\cer, 
J.~Rak\mpi, 
I.~Ravinovich\wei, 
P.~Rehak\bnl, 
M.~Richter\hei,
H.~Sako\gsi, 
W.~Schmitz\hei, 
S.~Sedykh\gsi,
W.~Seipp\hei,
A.~Sharma\gsi,
S.~Shimansky\dub, 
J.~Sl\'{\i}vov\'a\hei,
H.\,J.~Specht\hei, 
J.~Stachel\hei,
M.~\v{S}umbera\rez, 
H.~Tilsner\hei, 
I.~Tserruya\wei, 
J.\,P.~Wessels\gsi, 
T.~Wienold\hei, 
B.~Windelband\hei, 
J.\,P.~Wurm\mpi, 
W.~Xie\wei, 
S.~Yurevich\hei, 
V.~Yurevich\dub \\ 
(CERES/NA45 Collaboration)}

\affiliation{
\rez NPI ASCR, \v{R}e\v{z}, Czech Republic\\
\gsi GSI, 64220 Darmstadt, Germany\\
\hei \mbox{Physikalisches~Institut~der~Universit\"{a}t Heidelberg,~69120 Heidelberg,~Germany}\\
\dub JINR, Dubna, Russia\\
\wei Weizmann Institute, Rehovot 76100, Israel\\
\sun \mbox{State University of New York at Stony Brook, Stony Brook, New
York 11794, U.S.A.}\\
\cer CERN, 1211 Geneva 23, Switzerland\\
\bnl \mbox{Brookhaven National Laboratory, Upton, New York 11973, USA}\\
\mpi \mbox{Max-Planck-Institut f\"{u}r Kernphysik, 69117~Heidelberg,~Germany}\\}
 
\begin{abstract} 
We report on first measurements of low-mass electron-positron pairs in Pb-Au
collisions at the SPS beam energy of 40 AGeV. The observed pair yield
integrated over the range of invariant masses 0.2$<$m$\le$1
GeV/c$^{2}$ is enhanced over the expectation from neutral meson decays
by a factor of
5.9$\pm$1.5(stat.)$\pm$1.2(syst. data)$\pm$1.8(syst. meson decays),
somewhat larger than previously observed at the higher energy of 158 AGeV. The
results are discussed with reference to model calculations based on
$\pi^{+}\pi^{-}$$\rightarrow$$e^{+}e^{-}$ annihilation with a modified 
$\rho$-propagator. They may be linked to chiral symmetry restoration
and support the notion that the in-medium modifications of the $\rho$
are more driven by baryon density than by temperature.
\end{abstract}
\maketitle

According to finite temperature lattice QCD, strongly interacting
matter will, at sufficiently high energy densities, undergo a phase
transition from a state of hadronic constituents to quark matter, a
plasma of deconfined quarks and gluons. At the same time, chiral
symmetry, spontaneously broken in the hadronic world, will be
restored. High-energy nucleus-nucleus collisions provide the only way
to investigate this issue in the laboratory. Among the different observables
used for diagnostics of the hot and dense fireball formed in these
collisions, lepton pairs are
particularly attractive. In contrast to hadrons, they directly probe
the entire evolution of the fireball; the prompt emission 
after creation and the absence of any final state interaction
conserve the primary information within the limits imposed by the
space-time folding over the emission period.

The CERES/NA45 experiment at the CERN SPS is the
only dielectron spectrometer in the field of ultrarelativistic nuclear
collisions, focused on the measurement of electron-positron pairs in
the invariant mass range m$_{ee}$$\le$1 GeV/c$^{2}$. In a series of
systematic measurements, CERES has previously found pronounced
differences between proton-induced reactions like p-Be, p-Au at 450
GeV~\cite{Agakichiev:1995xb,Agakichiev:mv} and nucleus-nucleus
collisions like S-Au at 200 AGeV~\cite{Agakichiev:1995xb} and Pb-Au at 158
AGeV~\cite{Agakichiev:1997au,Lenkeit:1999xu}.
While the superposition of known 
electromagnetic decays of produced neutral mesons can successfully
account for the measured $e^{+}e^{-}$ mass spectra in the proton case, 
a strong excess of $e^{+}e^{-}$ pairs above the expectation from
meson decays has consistently been observed in the
nuclear case, amounting for 158 AGeV Pb-Au collisions to a factor of
2.4$\pm$0.2(stat.)$\pm$0.6(syst. data)$\pm$0.7(syst. meson decays) in the
mass interval 0.2$<$m$_{ee}$$\le$1 GeV/c$^{2}$(combined 1995/96
data~\cite{Agakichiev:1997au,Lenkeit:1999xu,Agakichiev:nn}).
The observation of this excess has initiated an enormous theoretical
activity (see~\cite{Rapp:1999ej,Brown:2001nh} for recent
reviews). There is a general consensus that one observes
direct thermal radiation from the fireball, dominated by two-pion annihilation
$\pi^{+}\pi^{-}$$\rightarrow$$\rho$$\rightarrow$$e^{+}e^{-}$ with an
intermediate $\rho$ vector meson. The $\rho$ is of particular
relevance (more so than the other light vector mesons $\omega$ and
$\phi$), due to its short lifetime of 1.3 fm/c and its direct link to
chiral symmetry restoration~\cite{Pisarski:mq}. Indeed, a proper description of
the observed dilepton excess does require a strong modification of the 
intermediate $\rho$ in the hot and dense medium. The two main theoretical
alternatives for this modification are (i) 'Brown-Rho scaling' with an
explicit connection to the medium dependence of the chiral
condensate~\cite{Brown:2001nh,Brown:kk,Hatsuda:1991ez,Li:1995qm}~,
reducing the mass of the $\rho$ below the vacuum value, and
(ii) calculations of the $\rho$ spectral density on the basis of
$\rho$-hadron interactions~\cite{Rapp:1999ej,Rapp:1995zy,Rapp:1999us},
spreading the width of the $\rho$ above the vacuum
value. Theoretically, the modifications are more sensitive to the
baryon density than to the temperature of the fireball.

To critically examine the relative importance of baryon density and
temperature, an experimental variation of these parameters
seemed mandatory.
In this Letter, we present first results~\cite{Damjanovic:2001qc} on
the production of $e^{+}e^{-}$ pairs 
for Pb-Au collisions at the SPS beam energy of 40 AGeV, where a
lower temperature and a higher baryon density (by a factor of 1.5)
are expected as compared
to 158 AGeV~\cite{Adamova:2002ff}. We indeed observe a strong excess of 
pairs again; the enhancement factor relative to the expectation from
meson decays is now found to be
5.9$\pm$1.5(stat.)$\pm$1.2(syst. data)$\pm$1.8(syst. decays), i.e. somewhat
greater than at the higher beam energy.

The CERES/NA45 experiment is optimized for the
measurement of $e^{+}e^{-}$ pairs in the invariant mass range of
tens of MeV/c$^{2}$ up to 1 GeV/c$^{2}$. The acceptance, azimuthally
symmetric, covers the
pseudorapidity region 2.11$\le$$\eta_{L}$$\le$2.64 in the laboratory
frame, i.e. a window of $\Delta$$\eta$=$0.53$ around mid-rapidity. A detailed
description of the experiment can be found
in~\cite{Baur:1993mz,Agakichiev:jc,Agakichiev:ka}.
Here we summarize the most relevant features.
Two Si drift detectors (SIDC1/2)~\cite{Agakichiev:jc},
located 10/14 cm downstream of a segmented Au target, provide an
angle measurement of the charged 
particles and reconstruction of the vertex;
at the same time, they serve as a powerful handle to
recognize photon conversion- and Dalitz electron-positron pairs with their
small opening angles. The basic
discrimination between the rare electrons and the abundant hadrons is
done with two Ring Imaging Cherenkov detectors
(RICH1/2)~\cite{Baur:1993mz,Agakichiev:jc}.
For the run periods 1999/2000, the spectrometer was upgraded by the addition of a
cylindrical Time Projection Chamber (TPC)~\cite{Agakichiev:ka} with a
radial electric drift field, operating
inside a new magnetic field configuration. As a consequence,
the mass resolution is potentially improved from
the 1995/96 value of 5.5$\%$ to about 2$\%$ in the region of the
$\rho/\omega$. Since
the original magnetic field between the RICH's is no longer
required, it is turned off, and the two detectors are now used as an integral unit,
resulting in an improved electron efficiency ($\sim$0.9) and rejection power.

The results reported in this Letter were obtained from the analysis of 
data taken during the SPS run period in the fall 
of 1999, the first on $e^{+}e^{-}$-pair production after
completion of the upgrade.
The centrality threshold, based on a
lower limit of the integrated signal of the SIDC1 detector
(proportional to the total number of hits),
was set to correspond to the 30$\%$ most central fraction of the
geometrical cross section (same as for the 1996 data). Since the new
TPC readout was still being commissioned, only about 8 million events
were taken. Related problems also limited the track efficiency of the
TPC to about 0.43 and the mass resolution actually achieved to about
6$\%$ in the region of the $\rho/\omega$ (inferred from an independent
analysis of $\Lambda$ and $K_{0}$ production).

The physics signal to be analyzed, $e^{+}e^{-}$ pairs
with masses m$_{ee}$$>$0.2 GeV/c$^{2}$, has an abundance of only
10$^{-5}$ both relative to hadrons and to photons. The experimental
challenges to minimize the associated background are therefore
extremely high. The electron track reconstruction and hadron
suppression proceed in the following way. First the event vertex is
reconstructed from 100 or more particle trajectories traversing both
SIDC's. Track segments are then formed in the SIDC's. Independently,
Cherenkov ring candidates are reconstructed according to a pattern
recognition algorithm based on a two-step Hough-transformation for
candidate search and a subsequent ring fitting
procedure~\cite{Baur:1993mz,Agakichiev:jc}.
Track segments from the SIDC's and
RICH's are then matched, defining electron track candidates, and these 
candidates are linked to the TPC and matched there to TPC track
segments, independently preselected by $dE/dx$ information
characteristic for relativistic electrons. With a RICH radiator
threshold of $\gamma_{th}$=32, more than 99$\%$ of all charged hadrons are
already rejected on the SIDC-RICH level. The remaining small
pion contamination, due to accidental matches with fake-ring
electron track candidates and to true matches with high-$p_{t}$ pions
with asymptotic ring radii,
is completely removed by the
(momentum-dependent) $dE/dx$-cut in the TPC, a considerable
improvement compared to the
1995/1996 analysis. Finally, the remaining electron and positron
tracks are combined into pairs, and the type of pair is defined.

The photon-related background is much more severe. Even with minimal
material in the experiment, photons converting in
the target and in SIDC1 together with $\pi^{0}$-Dalitz decays still
exceed the expected high-mass signal by a factor of about 10$^{3}$,
thus constituting the vast majority of the electron tracks. 
Although the characteristics are
different (small pair opening angles and very low masses m$_{ee}$$<$0.2
GeV/c$^{2}$), the limited track reconstruction efficiency and acceptance
lead to a {\em combinatorial high-mass background} for events in
which two or more of these low-mass pairs are only partially
reconstructed. This is {\em the} central problem
of the experiment. It is dealt with in the following
way~\cite{Damjanovic:2001qc}.
Since the inclusive electron $p_{t}$-spectra from low-mass pairs are
considerably softer than those
of pairs with m$_{ee}$$>$0.2 GeV/c$^{2}$, a reduction of the combinatorial 
background by more than a factor of 10 can be obtained by pairing only electron
tracks with transverse momenta p$_{t}$$>$0.2 GeV/c. Further, a global
selection on opening angles $\Theta_{ee}$$\ge$35 mrad is applied to all pairs with
complete tracks. Next, conversion- and Dalitz pairs
with opening angles $\Theta_{ee}$$<$10
mrad which are not recognized as separate tracks with
two individual electron rings in the
RICH detectors are rejected by a correlated double-$dE/dx$ cut in each 
of the two SIDC's~\cite{Agakichiev:jc}. Electron tracks which have a second
SIDC-RICH electron candidate track within 70 mrad are also
rejected. Finally, identified Dalitz pairs
(m$_{ee}$$\le$0.2 GeV/c$^2$) are excluded from further
combinatorics. Altogether, a rejection factor of more than 100 is obtained.

The remaining combinatorial background is determined by the number of
like-sign pairs. The net physics signal $S$ is then obtained by
subtracting the like-sign contribution from the total unlike-sign
pairs as $S= N_{+-}-2(N_{++} \cdot N_{--})^{1/2}$. The final sample
after all cuts consists of 532 unlike-sign, 283 like-sign and
249$\pm$29 net unlike-sign pairs (signal-to-background ratio of 1/1)
for masses m$_{ee}$$\le$0.2 GeV/c$^{2}$, and 1236 unlike-sign, 1051 like-sign
and 185$\pm$48 net unlike-sign pairs (signal-to-background ratio of
1/6) for masses 0.2$<$m$_{ee}$$\le$1.15 GeV/c$^{2}$.

The raw net yields require an {\em absolute} normalization and an {\em
absolute} correction for reconstruction efficiency for further
discussion; an acceptance correction is not done. The
normalization refers to the total number of events and the
average number $\langle$$N_{ch}$$\rangle$ of charged par\-ti\-cles per
event within the $e^{+}e^{-}$-pair ra\-pid\-ity ac\-cep\-tance. The charged
par\-ti\-cle ra\-pid\-ity density distribution is extracted from an
independent analysis of the SIDC track segments; for the present
centrality selection, one gets
$\langle$$N_{ch}$$\rangle$=$\langle$$dN_{ch}/d\eta$$\rangle$$\cdot$$\Delta\eta$=216$\cdot$0.53=115.
The pair reconstruction efficiency $\epsilon$ is determined by a Monte Carlo
procedure in which simulated lepton pairs from neutral meson decays
with known characteristics (see below) are embedded into real data and 
are then analyzed using the same software chain as used for the data; this
'overlay-MC' provides the highest degree of realism, in particular for 
the description of event background and instrumental deficiencies. A value of
$\langle$$1/\epsilon$$\rangle^{-1}$=0.036 is obtained as the weighted
average over the $dN_{ch}/d\eta$-distribution (the only
essential functional dependence), using as weights the uncorrected
differential pair yield for m$_{ee}$$\le$0.2GeV/c$^{2}$.
The primary reason for this low value
(compared to 0.11 in 1996) is the 1999 status of the TPC readout,
while the influence of the low-mass pair rejection cuts on the
efficiency is found to be minor. This is confirmed by the stability of
the two net pair samples with masses $\le$ and $>$ 0.2GeV/c$^{2}$
as a function of the rejection steps over a wide dynamical range of the
signal-to-background ratio~\cite{Damjanovic:2001qc}.
\begin{figure}[b!]
\begin{center}
\vspace*{-0.3cm}
\includegraphics*[width=8cm,height=7.cm]{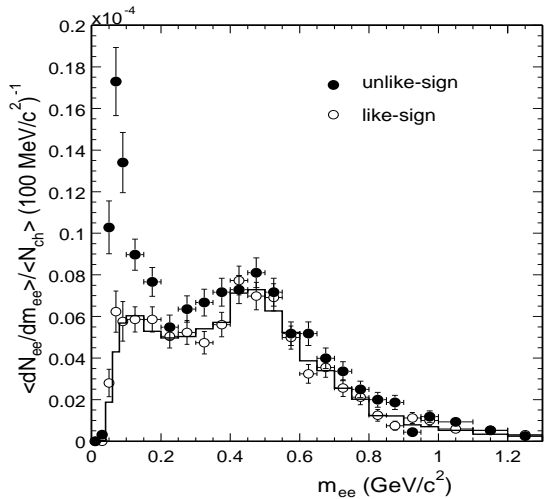}
\vspace*{-0.6cm}
   \caption{Mass spectra of unlike-sign and
like-sign pairs (circles). A high-statistics like-sign spectrum is shown 
as a histogram (see text). The peak at m$_{ee}$$>$0.4 GeV/c$^{2}$ is due to
the single-electron cut $p_{t}$$>$0.2 GeV/c.}
   \label{fig1}
\end{center}
\vspace*{-0.5cm}
\end{figure}

The normalized and efficiency-corrected mass spectra for the total
unlike-sign pairs and the like-sign pairs are shown in
Fig.~\ref{fig1}. The statistical bin-to-bin fluctuations of the
mass-differential {\em net} signal can be reduced by about a factor
of $\sqrt{2}$, if the {\em shape} of the like-sign combinatorial
background is determined with much better accuracy. Somewhat different 
from previous CERES analyses, this is done
here by using the like-sign combinatorial background after the first two
rejection steps (only $p_{t}$- and global $\Theta_{ee}$- cut) which contains
about 10 times more entries than the final background. This
'high-statistics' background is included in Fig.~\ref{fig1} as
a histogram, rescaled as to give the same integral as the final
background; the two are consistent within errors. The pair transverse
momentum spectra are treated similarly.

\begin{figure}[ht!]
\begin{center}
\includegraphics*[width=8.3 cm, height=7.3cm, bb=1 22 527 558]{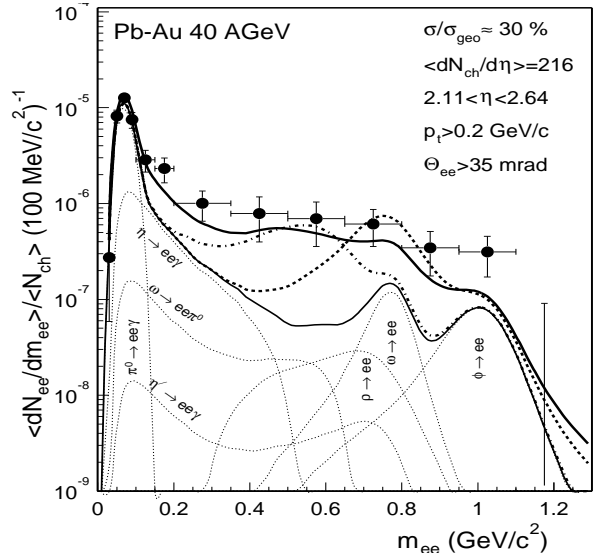}
\vspace*{-0.3cm}
\caption{Inclusive $e^{+}e^{-}$
mass spectrum, compared to the hadron decay cocktail (thin solid;
individual contributions thin dotted) and to theoretical model
calculations~\cite{Rapp:2002tw} based on $\pi^{+}\pi^{-}$ annihilation with an
unmodified $\rho$ (thick dashed), an in-medium dropping $\rho$ mass
(thick dashed-dotted) and an in-medium spread $\rho$ width (thick
solid). The model calculations contain the cocktail, but without the
$\rho$ to avoid double counting. The low-mass tail of the cocktail
$\rho$ is due to the inclusion of a $\pi^{+}\pi^{-}$ phase space
correction. The (weaker) tails of the $\omega$ and $\phi$ are caused
by electron bremsstrahlung.}
   \label{fig2}
\end{center}
\vspace*{-0.6cm}
\end{figure}

The resulting net invariant-mass spectrum of $e^{+}e^{-}$ pairs is
shown in Fig.~\ref{fig2}. The
errors attached to the data points are purely statistical. The
systematic errors contain contributions from the
normalization procedure
and the efficiency correction, amounting to an overall
error of 16$\%$ for masses m$_{ee}$$\le$0.2 GeV/c$^{2}$. In the high-mass
region m$_{ee}$$>$0.2 GeV/c$^{2}$, additional errors of about 20$\%$ arise
from remaining ambiguities in the rejection cuts and from the
statistical error of the background rescaling, raising here the
overall error to 26$\%$.
All values are quadratically small compared to the individual statistical errors
of the data points and are therefore not drawn. Integration of the
mass spectrum results in total normalized $e^{+}e^{-}$-pair yields per 
event per charged particle of
[7.6$\pm$0.8(stat.)$\pm$1.2(syst.)] $\cdot$10$^{-6}$ for m$_{ee}$$\le$0.2
GeV/c$^{2}$, and [5.6$\pm$1.4(stat.)$\pm$1.1(syst.)] $\cdot$10$^{-6}$ for
m$_{ee}$$>$0.2 GeV/c$^{2}$. The background rescaling used to improve
the differential spectrum has, of course, no relevance for the
integrals. Their statistical errors are those of the primary
samples quoted above, and the systematic error of the high-mass
integral m$_{ee}$$>$0.2 GeV/c$^{2}$ is only 20$\%$ (instead of 26$\%$). 

Fig.~\ref{fig2} also contains a comparison to the 
electromagnetic decays of the produced neutral mesons. The
evaluation of this 'hadron decay cocktail' follows our previous
procedure, described in detail for proton-induced reactions
in~\cite{Agakichiev:mv} and extended to the Pb-Au case
in~\cite{Lenkeit:1999xu,Agakichiev:nn}, including the
empirical systematics of hadron yields (statistical model values
where unmeasured) and of m$_{t}$-distributions (influenced by radial
flow). The simulations are subject to the same cuts as the data;
resolution broadening as inferred from hadron data (see above) and the 
effects of bremsstrahlung from the electrons traversing a total detector
material of 5.8$\%$ of a radiation length before the TPC are included.
\begin{figure}[b!]
\vspace*{-0.3cm}
{
\includegraphics*[width=4.7cm, height=6cm,clip=,bb= 32 0 477 531]{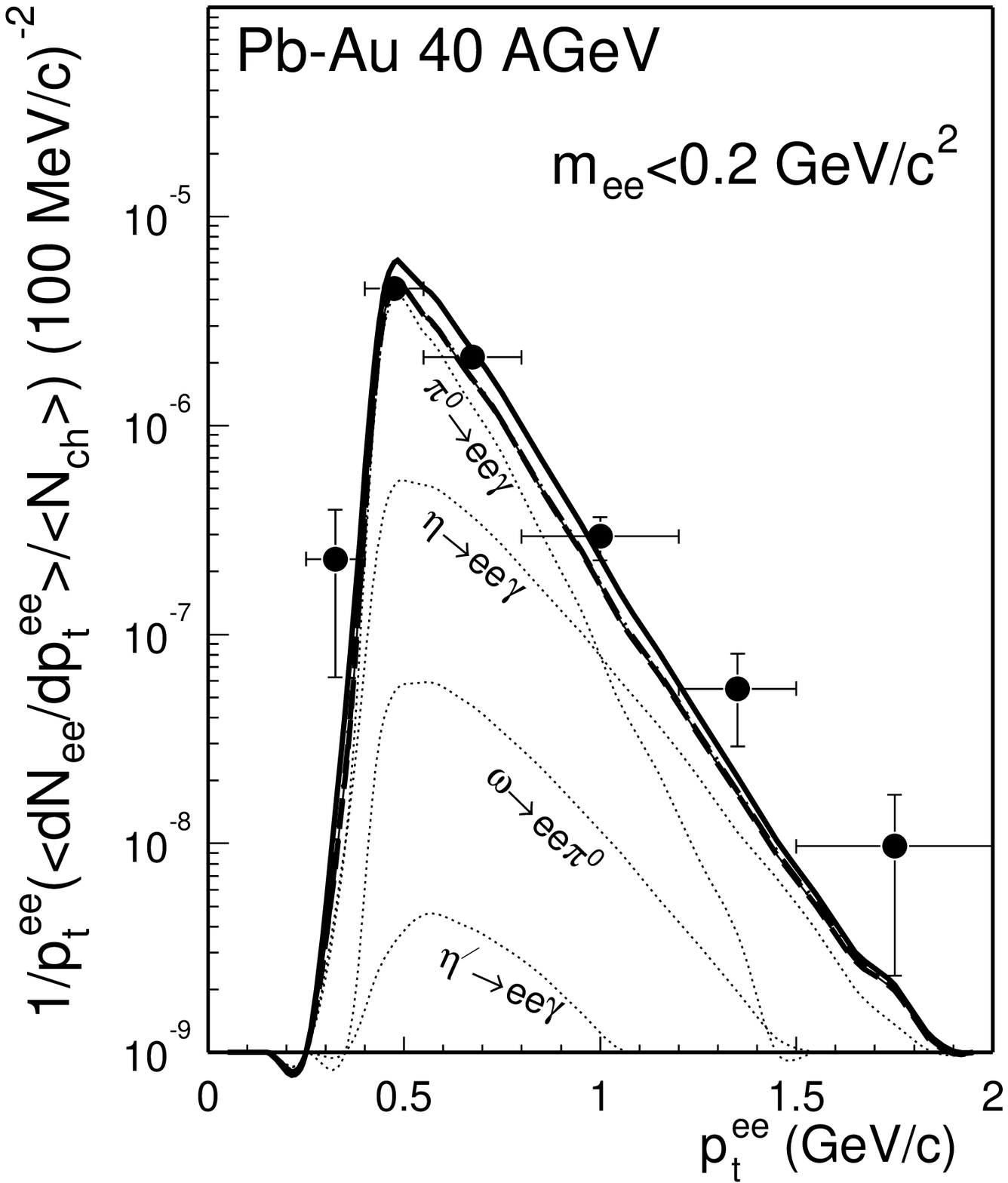}
\includegraphics*[width=3.8cm, height=6cm,clip=,bb= 118 0 477 531]{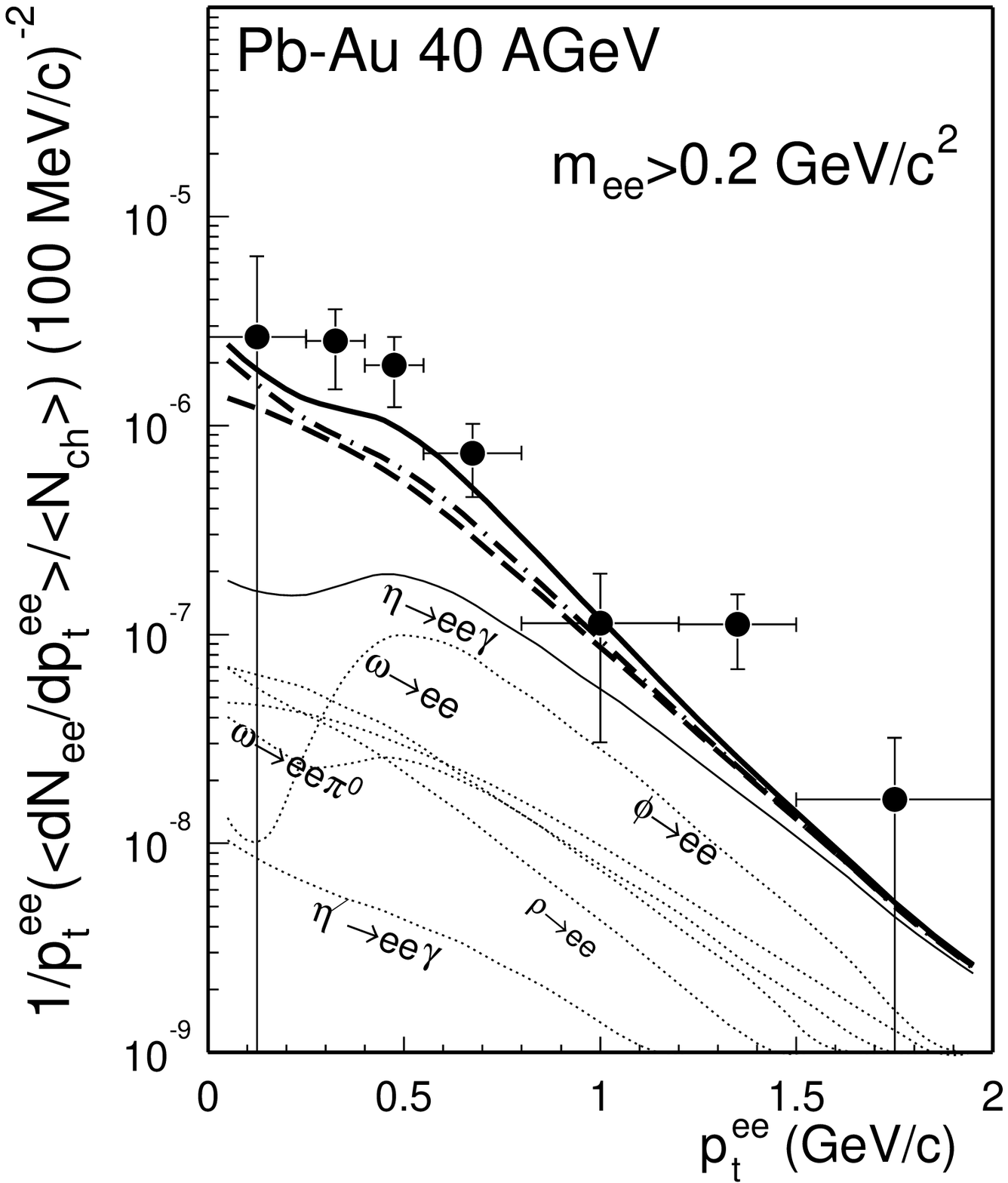}}
\caption{ Invariant
pair-$p_{t}^{ee}$ spectra for masses m$_{ee}$$\le$0.2 GeV/c$^{2}$ (left)
and m$_{ee}$$>$0.2 GeV/c$^{2}$ (right), compared to the hadron decay
cocktail and to theoretical model calculations~\cite{Rapp:2002tw}. The
cuts, labels and line codes are identical to those in Fig.~\ref{fig2}.}
   \label{fig3}
\end{figure}
For masses m$_{ee}$$\le$0.2 GeV/c$^{2}$,
the spectrum is dominated by $\pi^{0}$- and $\eta$ Dalitz decays, and
good agreement with a ratio data/decays of
1.18$\pm$0.13(stat.)$\pm$0.19(syst. data)$\pm$0.09(syst. decays) is
found. For masses m$_{ee}$$>$0.2 GeV/c$^{2}$, however, a structureless
continuum much above the hadron-decay level is seen, extending all
the way up to the region of the $\phi$. The enhancement factor
data/decays is
5.9$\pm$1.5(stat.)$\pm$1.2(syst. data)$\pm$1.8(syst. decays)
in this region, i.e. larger than the value at the higher beam energy of
158 AGeV with a statistical significance of 1.8$\sigma$ (data errors
added in quadrature); the systematic errors 'decays' of the decay
cocktail, dominated by uncertainties in the branching
ratios and transition form factors, essentially drop out in the comparison.

The invariant differential spectra in pair transverse momentum 
$p_{t}^{ee}$ are plotted in Fig.~\ref{fig3} separately for the two
mass regions m$_{ee}$$\le$ 0.2 and $>$0.2 GeV/c$^{2}$.
The steep cut-off in the distribution for $p_{t}^{ee}$$<$0.4 GeV/c
(left part) reflects the cut of the
single-electron $p_{t}$$\ge$0.2 GeV/c. Comparing data and hadron
decays, good agreement is seen
again for masses m$_{ee}$$\le$0.2
GeV/c$^{2}$, while the strong excess above the decay expectation
reappears for the high-mass region, mostly localized at low
$p_{t}^{ee}$$\le$0.6 GeV/c as observed before at 158
AGeV~\cite{Agakichiev:1997au,Lenkeit:1999xu,Agakichiev:nn}.

To illustrate most directly the relevance of the new results,
Figs.~\ref{fig2} and ~\ref{fig3} also contain the
{\em predictions} from theoretical model calculations~\cite{Rapp:2002tw}, based
on $\pi^{+}\pi^{-}$ annihilation as discussed in the introduction and
including the hadron decay contribution. The $\rho$-propagator is
treated in three ways - vacuum $\rho$, modifications through
'Brown-Rho scaling'~\cite{Brown:2001nh,Brown:kk,Hatsuda:1991ez,Li:1995qm},
and modifications through $\rho$-hadron
interactions~\cite{Rapp:1999ej,Rapp:1995zy,Rapp:1999us}. With
reference to the mass spectrum in Fig.~\ref{fig2}, the data clearly
rule out an unmodified $\rho$, but agree with the two in-medium
scenarios within errors (without being able to distinguish between
them). The pair transverse momentum spectra in Fig.~\ref{fig3} are also
reasonably well described, but the peculiar low-$p_{t}^{ee}$ rise (in
contrast to the shape of the hadron-decay spectra) appears here as a
basic feature of the pion-annihilation process without much
discrimination power towards in-medium effects. The integral yield of
the model calculations (thick solid line in Figs. 2 and 3) corresponds 
to a 'theoretical' enhancement factor of 4.2 at 40 AGeV, compared to
2.4 at the higher beam energy; the difference is consistent with the
observed trend.

We draw the following conclusions. The $e^{+}e^{-}$-pair yield for
m$>$0.2GeV/c$^{2}$ observed at 40 AGeV is enhanced over the
expectation from neutral meson decays. Compared to 158 AGeV, the
enhancement factor may even be larger, with a significance of
1.8$\sigma$. However, even if it would be the same, that would still 
be remarkable vis-a-vis the changes in conditions, i.e. less
multiplicity and less temperature at the lower energy. The only
quantity rising is the baryon density. Therefore, on very general
grounds and independent of the details of any model calculations, the sole existence of a strong
enhancement at 40 AGeV is consistent with the theoretical description of
$\pi^{+}\pi^{-}$ annihilation and a $\rho$ propagator dominantly
influenced by baryon density rather than
temperature~\cite{Rapp:1999ej,Brown:2001nh,Brown:kk,Hatsuda:1991ez,Li:1995qm,Rapp:1995zy,Rapp:1999us,Rapp:2002tw}.
It is also consistent with theoretical results that contributions from 
$q{\overline{q}}$ annihilation during the QGP phase are minor~\cite{Rapp:2002tw}.

\begin{acknowledgments}
This work was supported by the German BMBF, the U.S.~DoE, the
Israeli Science Foundation, and the MINERVA Foundation.
\end{acknowledgments}

\vspace*{-0.3cm}

 \end{document}